\documentstyle[12pt]{article}

\title{On the reaction field for interaction site \\
       models of polar systems}

\author{\em Igor~P.~OMELYAN \\ [1.5ex]
{\small \em Institute for Condensed Matter Physics,
            National Ukrainian Academy of Sciences} \\ [-8pt]
{\small \em 1~Svientsitsky St., UA-290011 Lviv, Ukraine \thanks
           {E-mail: nep@icmp.lviv.ua}} \\
\date{}}

\newcommand{\bms}[1]{\mbox{\boldmath $#1$}}
\newcommand{\bvs}[1]{\mbox{\footnotesize\boldmath $#1$}}
\newcommand{\scs}[1]{_{\stackrel{\ }{#1}}}

\begin{document}

\setlength{\abovedisplayskip}{18pt plus4pt minus6pt}
\setlength{\belowdisplayskip}{\abovedisplayskip}
\setlength{\abovedisplayshortskip}{12pt plus2pt minus4pt}
\setlength{\belowdisplayshortskip}{\abovedisplayshortskip}

\maketitle

\vspace{1cm}

\begin{abstract}

It is rigorously shown that the fluctuation formula, which is used
in simulations to calculate the dielectric constant of interaction
site models, corresponds to the reaction field with an individual
site cut-off rather than with the usual molecular center of mass
truncation. Within the molecular cut-off scheme, a modified reaction
field is proposed. An influence of the truncation effects is discussed
and examined by actual Monte Carlo simulations for a MCY water model.

\vspace{0.5cm}


\vspace{0.2cm}


\end{abstract}

\newpage

\section{Introduction}

\hspace{1em}  The calculation of dielectric quantities by computer
experiment requires an explicit consideration of effects associated
with the truncation of long-range interactions. The concrete success
in this direction has been achieved within the reaction field (RF)
geometry [1--5]. As a result, computer adapted dielectric theories
have been proposed [6--10]. In the framework of these theories, a bulk
dielectric constant can be determined on the basis of a fluctuation
formula via correlations obtained in simulations for finite samples.
However, main attention in the previous investigations has been focused
on polar systems with the point dipole interaction. As is now well
established, the model of point dipoles can not reproduce adequately
features of real polar liquids.

At the same time, attempts to apply the RF geometry for more realistic
interaction site (IS) models have also been made [11--13]. However, acting
within a semiphenomenological approach, it was not understood how to perform
the truncation of intermolecular potentials. As a consequence, the molecular
cut-off and the usual point dipole RF (PDRF) have been assumed. Obviously,
such an approach includes effects connected with finiteness of the molecule
inconsistently. Indeed, the interdipolar potential is replaced by site-site
Coulomb interactions, whereas the RF is remained in its usual form. An
additional complication for IS models consists in a spatial distribution
of charges and this fact is not taken into account by the standard PDRF
geometry.

In the present paper we propose two alternative approaches to remedy this
situation. The first one follows from the usual fluctuation formula which
is constructed, however, on the microscopic operator of polarization density
for IS models. This leads to an ISRF geometry, where the cut-off radius is
applied with respect to individual charges rather than to the molecule as a
whole. Nevertheless, the molecular cut-off scheme can also be acceptable,
but the reaction field together with the fluctuation formula need to be
corrected. In the second approach a molecular RF (MRF) geometry is proposed
and a new quadrupole term is identified. On the basis of a MCY water model
we show that uncertainties of the dielectric quantities can be significant
if the standard PDRF geometry is used in computer simulations.

\vspace{12pt}

\section{Interaction site reaction field}

\hspace{1em}  We consider an isotropic, classical system of $N$ identical
molecules enclosed in volume $V$. The microscopic electrostatic field
created by the molecules at point $\bms{r} \in V$ is equal to
\begin{equation}
\bms{\hat E}(\bms{r}) = \sum \limits_{i=1}^N \sum \limits_a q \scs{a}
\frac{\bms{r}-\bms{r}_i^a}{|\bms{r}-\bms{r}_i^a|^3} = \int \limits_V
\bms{L}(\bms{r}-\bms{r}') \hat{Q}(\bms{r}') {\rm d} \bms{r}' \ ,
\end{equation}
where $\bms{r}_i^a$ denotes the position for charge $q \scs{a}$ of $i$th
molecule, $\hat Q(\bms{r})=\sum_{i, a} q \scs{a} \delta(\bms{r}-\bms{r}_i^a)$
is the microscopic operator of charge density, $\bms{L}(\bms{\rho})=-\bms
{\nabla} \ 1/\rho$ and the summation extends over all molecules and charged
sites. For the investigation of dielectric properties, it is more convenient
to rewrite the electric field (1) in the polarization representation
\begin{equation}
\bms{\hat E}(\bms{r}) = \int \limits_V \bms{\rm T}(\bms{r}-\bms{r}')
\bms{\hat P}(\bms{r}') {\rm d} \bms{r}' = - \frac{4\pi}{3} \bms{\hat P}
(\bms{r}) + \lim_{\rho \to +0} \!\! \mathop{\int \limits_V} \limits_
{\rho<|\bvs{r}-\bvs{r}'|} \relax \!\!\!\! \bms{\rm T}(\bms{r}-\bms{r}')
\bms{\hat P}(\bms{r}') {\rm d} \bms{r}' \ .
\end{equation}
Here $\bms{\rm T}(\bms{\rho}) = \bms{\nabla} \bms{\nabla} \ 1/\rho$ is
the dipole-dipole tensor, $\bms{\hat P}(\bms{r})$ denotes the microscopic
operator of polarization density, defined as $\bms{\nabla} \bms{\cdot}
\bms{\hat P}(\bms{r})=-\hat Q(\bms{r})$, and the singularity $\lim_{\rho
\to 0} \bms{\rm T}(\bms{\rho})= -4\pi /3 \ \delta(\bms{\rho}) \bms{\rm I}$
has been avoided, where $\bms {\rm I}$ is the unit tensor of the second rank.
The both charge (1) and polarization (2) representations are equivalent and
applicable for infinite ($N, V \to \infty$) systems.

In simulations, which deal with finite samples, the sum (1) can not be
calculated exactly taking into account an infinitely large number of terms.
Therefore, we must restrict ourselves to a finite set of terms in (1) or
to a finite range of the integration in (1) and (2) for which $|\bms{r}-
\bms{r}'| \le R$, where $R$ is a cut-off radius. Now the following problem
appears. How to estimate the cut-off field caused by the integration over
unaccessible region $|\bms{r}-\bms{r}'| > R$? The solution of this problem
has been found for the first time for systems with point dipoles in the RF
geometry. The result for conducting boundary conditions is [7, 8]
\begin{equation}
\bms{\hat E}(\bms{r}) \approx \bms{\hat E} \scs{R}^{\rm RF}(\bms{r}) =
- \frac{4\pi}{3} \bms{\hat P}(\bms{r}) + \lim_{\rho \to +0} \!\!
\mathop{\int \limits_{V,\ \rm tbc}} \limits_{\rho<|\bvs{r}-\bvs{r}'|\le R}
\relax \!\!\!\! \bigg( \bms{\rm T}(\bms{r}-\bms{r}') +
\frac{\bms{\rm I}}{R^3} \bigg) \bms{\hat P}(\bms{r}')
{\rm d} \bms{r}' \ ,
\end{equation}
where a cubic finite sample and toroidal boundary conditions (TBC) have been
used, so that $R \le \sqrt[3]{V}/2$. The additional term $\bms{\rm I}/R^3$
in the right-hand site of (3) describes the RF which is used for an
approximation of the real cut-off field. For a pure spherical cut-off (SC)
without the RF correction, we have $\bms{\hat E} \scs{R}^{\rm SC}(\bms{r}) =
\displaystyle \int \gamma (|\bms{r}-\bms{r}'|) \bms{L}(\bms{r}-\bms{r}')
\hat{Q}(\bms{r}') {\rm d} \bms{r}'$, where $\gamma(\rho)=1$ if $\rho
\le R$ and $\gamma (\rho)=0$ otherwise. Obviously, that $\lim_{R \to
\infty} \bms{\hat E} \scs{R}^{\rm SC}(\bms{r})=\lim_{R \to \infty} \bms
{\hat E} \scs{R}^{\rm RF}(\bms{r})=\bms{\hat E}(\bms{r})$.

Let us perform the spatial Fourier transform ${\cal F} (\bms{k}) = \int {\rm
d} \bms{r} {\mbox{\large e}}^{-{\rm i}\bvs{k\!\cdot\!r}} \ {\cal F}(\bms{r})$
for arbitrary functions ${\cal F}$. Then one obtains
\vspace{-1mm}
\begin{equation}
\bms{\hat E}\scs{R}^{\rm SC}(\bms{k}) = \bms{L}(\bms{k}) \hat{Q}(\bms{k}) \ ,
\ \ \ \ \  \bms{\hat E} \scs{R}^{\rm RF}(\bms{k}) = - \frac{4\pi}{3}
\bms{\hat P}(\bms{k}) + \Big( \bms{\rm T}(\bms{k}) + 4\pi \frac{j\scs{1}(kR)}
{kR} \bms{\rm I} \Big) \bms{\hat P}(\bms{k}) \ , \\ [-2mm]
\end{equation}
where
\vspace{-3mm}
\begin{equation}
\bms{L}(\bms{k}) = - 4\pi \Big( 1-j\scs{0}(kR) \Big) \frac{{\rm i}
\bms{k}}{k^2} \ , \ \ \ \ \ \bms{\rm T}(\bms{k}) = - \frac{4\pi}{3}
\Big( 1-3 \frac{j\scs{1}(kR)}{kR} \Big) \left( 3 \bms{\hat k}\bms{\hat k}-
\bms{\rm I} \right) \ , \ \ \\ [1mm]
\end{equation}
$\hat Q(\bms{k})=\sum_{i, a} q \scs{a} {\mbox{\large e}}^{-{\rm i} \bvs{k\!
\cdot\!r}_i^a}=-{\rm i} \bms{k} \bms{\cdot} \bms{\hat P} (\bms{k})$, \
$\bms{k}=2 \pi \bms{n}/\sqrt[3]{V}$ is one of the allowed wavevectors
of the reciprocal lattice, $\bms{n}$ designates a vector with integer
components, $k=|\bms{k}|$, $\bms{\hat k}=\bms{k}/k$ and $j\scs{0}(z)=
\sin(z)/z$, $j\scs{1}(z)=- \cos(z)/z+\sin(z)/z^2$ are the spherical Bessel
functions of zero and first order, respectively. In view of (5), the
relations (4) transform into
\begin{equation}
\bms{\hat E} \scs{R}^{\rm SC}(\bms{k}) = -4\pi \Big( 1 - j\scs{0}(kR) \Big)
\bms{\hat P}_{\rm L}(\bms{k}) \ , \ \ \ \ \ \bms{\hat E} \scs{R}^{\rm RF}
(\bms{k}) = -4\pi \Big( 1-3 \frac{j\scs{1}(kR)}{kR} \Big)
\bms{\hat P}_{\rm L}(\bms{k}) \ ,
\end{equation}
where $\bms{\hat P}_{\rm L}(\bms{k})=\bms{\hat k} \bms{\hat k} \bms{\cdot}
\bms{\hat P}(\bms{k})={\rm i}\bms{k}\hat Q(\bms{k})/k^2$ is the longitudinal
component of the microscopic operator of polarization density.

It is easy to see from (6) that the both functions $\bms{\hat E} \scs{R}^{\rm
SC}(\bms{k})$ and $\bms{\hat E} \scs{R}^{\rm RF}(\bms{k})$ tend to the same
value $\bms{\hat E}(\bms{k})=-4\pi \bms{\hat P}_{\rm L}(\bms{k})$ of the
infinite system at $R \to \infty$ ($k \ne 0$). However, the results converge
as $R^{-1}$ for the pure SC scheme, while as $R^{-2}$ in the RF geometry,
i.e., more quickly, because a main part of the truncation effects is taken
into account by the RF. This is very important in our case, where we hope
to reproduce features of infinite systems on the basis of finite samples.
That is why the pure truncation, which is standard for simple fluids with
short-range potentials, is generally not recommended for polar systems
with long-range nature of the dipolar interaction. The influence of the
TBC and the difference between micro- and canonical ensembles are of order
$N^{-1} \sim R^{-3}$ [14] and, therefore, they can be excluded from our
consideration. It is worth mentioning that electrostatic fields are pure
longitudinal. They can be defined via the longitudinal component of the
microscopic operator of polarization density, that is confirmed by Eq.~(6).

Let us enclose the system in an external electrostatic field $\bms{E} \scs
{0}(\bms{r})$. The material relation between the macroscopic polarization
$\bms{P}_{\rm L}(\bms{k}) = \left< \bms{\hat P}_{\rm L}(\bms{k}) \right>$ in
the weak external field and total macroscopic field is $4\pi \bms{P}_{\rm L}
(\bms{k}) = \Big(\varepsilon \scs{\rm L}(k) - 1 \Big) \bms{E}_{\rm L}
(\bms{k})$, where $\varepsilon \scs{\rm L}(k)$ denotes the longitudinal
wavevector-dependent dielectric constant. Applying the first-order
perturbation theory with respect to $\bms{E} \scs{0}$ yields for rigid
molecules $V k_{\rm B} T \bms{P}_{\rm L}(\bms{k}) = \left< \bms{\hat P}_{\rm
L}(\bms{k}) \bms{\cdot} \bms{\hat P}_{\rm L}(-\bms{k}) \right> \scs{0}
\bms{E} \scs{0}(\bms{k})$, where $k_{\rm B}$ and $T$ are Boltzmann's constant
and temperature, respectively, and $\left< ... \right> \scs{0}$ is the
equilibrium average in the absence of the external field. Then, taking into
account that $\bms{E}_{\rm L}(\bms{k})=\bms{E} \scs{0} (\bms{k})+\left<
\bms{\hat E} \scs{R}^{\rm RF}(\bms{k}) \right>$ and eliminating $\bms{E}
\scs{0}(\bms{k})$, we obtain the fluctuation formula
\begin{equation}
\frac{\varepsilon \scs{\rm L}(k) - 1}{\varepsilon \scs{\rm L}(k)}=
\frac{9y G \scs{\rm L}(k)}{1+27y G \scs{\rm L}(k) j\scs{1}(kR)/(kR)} =
9y g \scs{\rm L}(k) \ .
\end{equation}
Here $G \scs{\rm L}(k)= \left<\bms{\hat P}_{\rm L}(\bms{k}) \bms{\cdot}
\bms{\hat P}_{\rm L}(-\bms{k}) \right> \scs{0} \Big/ N \mu^2$ is the
longitudinal component of the finite-system wavevector-dependent Kirkwood
factor, $y=4\pi N \mu^2 \Big/ 9Vk_{\rm B}T$ and $\mu=|\bms{\mu}_i|=|\sum_a
q \scs{a} \bms{r}_i^a|$ denotes the permanent magnitude of molecule's dipole
moment. It is necessary to note that we consider rigid IS molecules so that
effects associated with molecular and electronic polarizabilities are not
included in our investigation. In the case of $R \to \infty$, we have
$j\scs{1}(kR)/(kR) \to 0$ and computer adapted formula (7) reduces to the
well-known fluctuation formula for macroscopic systems in terms of the
infinite-system Kirkwood factor $g \scs{\rm L}(k)=\lim_{R \to \infty}
G \scs{\rm L}(k)$.

As was mentioned earlier, the electric field $\bms{\hat E} \scs{R}^{\rm RF}$
in the form (3), (4) as well as the fluctuation formula (7) have been
proposed for the first time to investigate polar systems of point dipoles
[8]. However, acting within a semiphenomenological framework, it was not
understood how to perform the truncation of the intermolecular potential
$\varphi \scs{ij}$ at attempts to extend this formula for IS models. As a
result, the molecular cut-off $r_{ij}=|\bms{r}_i-\bms{r}_j| \le R$, where
$\bms{r}_i$ is the center of mass for $i$th molecule, and the usual
PDRF have been suggested [11--13]:
\vspace{-3.5mm}
\begin{equation}
\varphi \scs{ij} = \sum_{a,b} \frac{q \scs{a} q \scs{b}}
{|\bms{r}_i^a-\bms{r}_j^b|} - \frac{\bms{\mu}_i \bms{\cdot}
\bms{\mu}_j}{R^3} \ , \ \ \ \ \ r_{ij} \le R \ . \\ [1mm]
\end{equation}

It is essentially to emphasize that the fluctuation formula (7) takes into
account finiteness of the system explicitly by the factor $j\scs{1}(kR)/
(kR)$. As a result, if the system size is sufficiently large (terms of order
$R^{-2}$ can be neglected), the bulk ($N,V \to \infty$) dielectric constant
can be reproduced via the finite-system Kirkwood factor $G \scs{\rm L}(k)$
which depends on $R$ in a characteristic way. However, to achieve this
self-consistency in the evaluation of the bulk dielectric constant, the
equilibrium averaging in $G \scs{\rm L}(k)$ must be calculated for systems
with the intermolecular potential which leads exactly to the microscopic
electric field $\bms{\hat E} \scs{R}^{\rm RF}(\bms{r})$ (3). As we shall
below, the intermolecular potential (8) does not obey this condition.

To derive the exact intermolecular potential in the charge representation, we
perform the inverse Fourier transform $\bms{\hat E} \scs{R}^{\rm RF}(\bms{r})
=\frac{1}{(2\pi)^3} \displaystyle \int {\rm d} \bms{k} \bms{\hat E} \scs{R}^
{\rm RF}(\bms{k}) {\mbox{\large e}}^{{\rm i}\bvs{k\!\cdot\!r}}$ and obtain
using (6)
\begin{equation}
\bms{\hat E} \scs{R}^{\rm RF}(\bms{r}) = \sum_{i,a} q \scs{a} \frac{\bms{r}-
\bms{r}_i^a}{|\bms{r}-\bms{r}_i^a|^3} \left(1-\frac{6}{\pi} \frac{|\bms{r}-
\bms{r}_i^a|^2}{R} \int \limits_0^\infty j\scs{1}(kR) j\scs{1}(k|\bms{r}-
\bms{r}_i^a|) {\rm d} k \right) \ .
\end{equation}
Taking into account that $\frac{6}{\pi} \int_0^\infty j\scs{1}(kR) j\scs{1}
(k \rho) {\rm d} k= \rho/R^2$ if $\rho \le R$ and is equal to $R/\rho^2$ if
$\rho > R$, we have
\begin{equation}
\bms{\hat E} \scs{R}^{\rm RF}(\bms{r}) = \sum_{i,a} q \scs{a} \frac{\bms{r}-
\bms{r}_i^a}{|\bms{r}-\bms{r}_i^a|^3} \left( 1 - \frac{|\bms{r}-
\bms{r}_i^a|^3}{R^3} \right) \ \ \ {\rm if} \ \ |\bms{r}-\bms{r}_i^a| \le R
\end{equation}
and $\bms{\hat E} \scs{R}^{\rm RF}(\bms{r})=0$ otherwise, where the first
term in the right-hand side is the Coulomb field, while the second
contribution corresponds to the RF in the IS description.

In order to understand nature of this field, we consider a spherical cavity
of radius $R$ with the center at point $\bms{r}$, embedded in an infinite
conducting medium. Let us place a point charge $q \scs{a}$ at point $\bms
{r}_i^a$ in the cavity, so that $|\bms{r}-\bms{r}_i^a| \le R$. The total
electric field $\bms{e}_i^a(\bms{r})$ at point $\bms{r}$ consists of the
field due to the charge $q \scs{a}$ and the field created by induced
charges located on the surface of the cavity. According to the method of
electrostatic images [5], this last field can be presented as the field of
an imaginary charge $q^* \scs{a}=-q \scs{a} R/|\bms{r}-\bms{r}_i^a|$ which
is located at point ${\bms{r}^*}^a_i=\bms{r}-R^2 (\bms{r}-\bms{r}_i^a)/
|\bms{r}-\bms{r}_i^a|^2$ outside the sphere. Then $\bms{e}_i^a(\bms{r})=
q \scs{a} (\bms{r}-\bms{r}_i^a)/|\bms{r}-\bms{r}_i^a|^3+q^* \scs{a} (\bms
{r}-{\bms{r}^*}^a_i)/|\bms{r}-{\bms{r}^*}^a_i|^3=q \scs{a} (\bms{r}-
\bms{r}_i^a) (1/|\bms{r}-\bms{r}_i^a|^3-1/R^3)$ that is completely in
line with the term of sum (10).

In the potential representation ($\bms{\hat E} \scs{R}^{\rm RF}(\bms{r})= -
\bms{\nabla} {\mit \Phi}(\bms{r})$), we obtain ${\mit \Phi}(\bms{r})=
\sum_{i,a} \phi_i^a(\bms{r})$, where $\phi_i^a(\bms{r}) = q \scs{a} \
(1/\rho_i^a + \frac12 {\rho_i^a}^2/R^3+C)$, $\rho_i^a=|\bms{r}-
\bms{r}_i^a|$ and $C$ is, in general, an arbitrary constant which for
infinite systems is chosen as $\phi_i^a |_{\rho_i^a \to \infty}=0$.
In our case, according to the toroidal boundary conventional, $\phi_i^a
|_{\rho_i^a = R}=0$ whence $C=-3/2 \, R^{-1}$. Then the intermolecular
potential of interaction is $\varphi \scs{ij}= \sum_{a,b} q \scs{b}
\phi_i^a(\bms{r}_j^b) = \sum_{a,b} q \scs{a} \phi_j^b(\bms{r}_i^a)=
\sum_{a,b} \varphi \scs{ij}^{ab}$, where
\begin{equation}
\varphi \scs{ij}^{ab} = \left \{
\begin{array}{clc}
\displaystyle q \scs{a} q \scs{b} \left( \frac{1}{|\bms{r}_i^a-
\bms{r}_j^b|} + \frac12 \frac{|\bms{r}_i^a-\bms{r}_j^b|^2}{R^3} - \frac{3}
{2R} \right) \! & , & \ \ |\bms{r}_i^a-\bms{r}_j^b| \le R \\ [4mm]
0 \! & , & \ \ |\bms{r}_i^a-\bms{r}_j^b| > R
\end{array}
\right.
\end{equation}
and the site-site cut-off is performed.

It is easily seen from (11) that the ISRF part $\frac12 \sum_{a,b} q \scs{a}
q \scs{b} |\bms{r}_i^a-\bms{r}_j^b|^2/R^3$ transforms into the usual form
$-\bms{\mu}_i \bms{\cdot} \bms{\mu}_j/R^3$ of point dipoles for $r_{ij} \le
R-d$ only, where $d=2 \max|\bms{\delta}_i^a |$ is the diameter of the
molecule and $\bms{\delta}_i^a =\bms{r}_i^a-\bms{r}_i$. In the case if the
molecular rather than the site-site cut-off is applied to the potential
(11), this transformation is valid for arbitrary $r_{ij} \le R$. Moreover,
in the last case the constant $C=-3/2 \, R^{-1}$ is canceled owing
electroneutrality ($\sum_a q \scs{a}=0$) of the molecule and we recover
the result (8) of previous work [11]. However, the potential of interaction
(11) corresponds completely to the conditions at which the fluctuation
formula (7) is derived. Therefore, this potential, instead of (8), must be
used in simulations to obtain a correct value for the dielectric constant.

\vspace{12pt}

\section{Molecular reaction field}

\hspace{1em}  In the case of point dipoles, where $d \to +0$, $q \scs{a}
\to \infty$ provided $\mu \to const$, both (8) and (11) representations
are identical and reduced to the well-known result
\vspace{-1.5mm}
\begin{equation}
\varphi \scs{ij} = - \bms{\mu}_{i} \bms{\cdot} \bms{\rm T}(\bms{r}_{ij})
\bms{\cdot} \bms{\mu}_{j} - \frac{\bms{\mu}_i \bms{\cdot} \bms{\mu}_j}
{R^3} \ , \ \ \ \ \ r_{ij} \le R \\ [-1.5mm]
\end{equation}
for the interdipolar interaction in the RF geometry. It is easy to see that
in the case of IS models, the intermolecular potential (8) takes into
account effects associated with finiteness of the molecule inconsistently.
For example, the interdipolar potential is replaced by the real site-site
Coulomb ones, whereas the reaction field is remained in its usual form of
point dipoles. From this point of view a natural question of how to improve
the RF within the molecular cut-off scheme arises. The simplest way to solve
this problem lies in the following.

Let us consider the mentioned above spherical cavity, centered now at some
fixed point $\bms{r} \scs{0}$, in the infinite conducting medium. We place
an $i$th molecule in such a way that all sites of the molecule would be
located in the cavity. This condition is fulfilled providing $|\bms{r}_i -
\bms{r} \scs{0}| \le R_d \equiv R-d/2$. The potential of a molecular reaction
field at point $\bms{r}$ belonging the cavity can be presented, according to
the method of electrostatic images, as
\begin{equation}
\varphi^{\rm RF}_i(\bms{r}) = \sum_a \frac{q^* \scs{a}}{|\bms{\rho}-
{\bms{\rho}^*}_i^a|} = - \sum_a \frac{q \scs{a} R/\rho_i^a} {\bigg|
\displaystyle \bms{\rho} - \bigg(\frac{R}{\rho_i^a}\bigg)^2
\bms{\rho}_i^a \bigg|} = - \sum_a \frac{q \scs{a}}{\bigg| \displaystyle
\frac{\rho_i^a}{R} \bms{\rho} - \frac{R}{\rho_i^a} \bms{\rho}_i^a \bigg|} \ ,
\end{equation}
where $\bms{\rho}=\bms{r}-\bms{r} \scs{0}$ and $\bms{\rho}_i^a=\bms{r}_i^a-
\bms{r} \scs{0}$. Differentiating (13) over $\bms{r}$ at point $\bms{r}
\scs{0}$ yields
\begin{equation}
\frac{\partial \varphi^{\rm RF}_i(\bms{r})}{\partial \bms{r}} \bigg|_
{\bms{r}_0} = - \frac{\bms{\mu}_i}{R^3} \ , \ \ \ \ \ \frac{\partial^2
\varphi^{\rm RF}_i(\bms{r})}{\partial \bms{r} \partial \bms{r}} \bigg|_
{\bms{r}_0} = - \frac{\bms{\rm q}_i^{\bvs{r}_0}}{R^5} \ , \ \ \ \ \
\frac{\partial^3 \varphi^{\rm RF}_i(\bms{r})}{\partial \bms{r} \partial
\bms{r} \partial \bms{r}} \bigg|_{\bms{r}_0} = -
\frac{\bms{\rm g}_i^{\bvs{r}_0}}{R^7} \ , \ \ \ldots
\end{equation}
Here $\bms{\mu}_i=\sum_a q \scs{a} \bms{\rho}_i^a=\sum_a q \scs{a} \bms
{\delta}_i^a$ is the dipole moment of $i$th molecule, which does not depend
on $\bms{r} \scs{0}$ owing electroneutrality of the molecule, while $\bms
{\rm q}_i^{\bvs{r}_0}=\sum_a q \scs{a} (3 \bms{\rho}_i^a \bms{\rho}_i^a -
{\rho_i^a}^2 \bms{\rm I})$ and $\bms{\rm g}_i^{\bvs{r}_0}$ are the tensors
of quadrupole and octupole moments, correspondingly, of $i$th molecule with
respect to $\bms{r} \scs{0}$. The third rank tensor $\bms{\rm g}_i^{\bvs
{r}_0}$ has the following components ${\bms{\rm g}_i^{\bvs{r}_0}}_{\!\!
\alpha \beta \gamma}=3 \sum_a q \scs{a} \Big( 5 {\bms{\rho}_i^a}_\alpha
{\bms{\rho}_i^a}_\beta {\bms{\rho}_i^a}_\gamma - {\rho_i^a}^2 ({\bms
{\rho}_i^a}_\alpha \delta_{\beta \gamma} + {\bms{\rho}_i^a}_\beta
\delta_{\alpha \gamma} + {\bms{\rho}_i^a}_\gamma \delta_{\alpha \beta})
\Big)$. It is more convenient to present multipoles of higher order
with respect to the molecular center of mass. For the tensor of quadrupole
moment we obtain $\bms{\rm q}_i^{\bvs{r}_0}=\bms{\rm q}_i+\bms{\rm w}_i$,
where $\bms{\rm q}_i=\sum_a q \scs{a} (3 \bms{\delta}_i^a \bms{\delta}_i^a -
{{\delta}_i^a}^2 \bms{\rm I})$ is the tensor of quadrupole moment of $i$th
molecule with respect to its center of mass, $\bms{\rm w}_i=3(\bms{\mu}_i
\bms{\rho}_i + \bms{\rho}_i \bms{\mu}_i) - 2 \bms{\mu}_i \bms{\cdot} \bms
{\rho}_i \bms{\rm I}$ and  $\bms{\rho}_i=\bms{r}_i-\bms{r} \scs{0}$. It is
necessary to underline that tensor $\bms{\rm q}_i$ is split into dynamical
$\bms{\omega}_i=\sum_a q \scs{a} \bms{\delta}_i^a \bms{\delta}_i^a$ and
conservative $\sum_a q \scs{a} {{\delta}_i^a}^2 \bms{\rm I}$ parts for
rigid molecules.

Putting $\bms{r} \scs{0}=\bms{r}_j$ and assuming $d \ll R$, we obtain the
energy of $j$th molecule in the MRF of $i$th molecule
\begin{equation}
\phi^{\rm RF}_{ji} = \bms{\mu}_j \bms{\cdot} \frac{\partial
\varphi^{\rm RF}_i(\bms{r})}{\partial \bms{r}} \bigg|_
{\bms{r}_j} + \frac16 \bms{\rm q}_j \bms{:} \frac{\partial^2
\varphi^{\rm RF}_i(\bms{r})}{\partial \bms{r} \partial \bms{r}}
\bigg|_{\bms{r}_j} + ... = - \frac{\bms{\mu}_j \bms{\cdot} \bms
{\mu}_i}{R^3} - \frac16 \frac{\bms{\rm q}_j \bms{:}
\bms{\rm q}_i^{\bvs{r}_j}}{R^5} + ... \ ,
\end{equation}
where multipoles of higher order have been neglected. Finally, using the
RF potential $\varphi^{\rm RF}_{ij}=(\phi^{\rm RF}_{ij}+\phi^{\rm RF}_{ji})
/2$ yields the desired intermolecular potential
\begin{equation}
\varphi \scs{ij} = \left \{
\begin{array}{clc}
\displaystyle \sum_{a,b} \frac{q \scs{a} q \scs{b}} {|\bms{r}_i^a-
\bms{r}_j^b|} - \frac{\bms{\mu}_i \bms{\cdot} \bms{\mu}_j}{R^3} -
\frac {\bms{\rm q}_i \bms{:} \bms{\rm q}_j - 3 (\bms{\rm q}_i \bms{:}
\bms{\mu}_j \bms{r}_{ij} + \bms{\rm q}_j \bms{:} \bms{\mu}_i \bms{r}_{ji})}
{6R^5} \! & , & \ \ r_{ij} \le R_d \\ [4mm] 0 \! & , & \ \ r_{ij} > R_d
\end{array}
\right. \\
\end{equation}
where equality $\bms{\rm q} \bms{:} \bms{\rm I}=0$ has been used.

The total reaction field, created by all molecules at point $\bms{r}$ near
$\bms{r} \scs{0}$ is
\vspace{0pt}
\begin{equation}
\bms{E} \scs{\rm RF} (\bms{r}) = - \sum_i^{\rho_i \le R_d}
\frac{\partial \varphi^{\rm RF}_i(\bms{r})}{\partial \bms{r}} =
\frac{\bms{M}(R_d)}{R^3} +\frac{\bms{\rm Q}(R_d) + \bms{\rm W}(R_d)}
{R^5} \bms{\rho} + \ldots \ ,
\end{equation}
where $\bms{M}(R_d)=\sum_i^{\rho_i \le R_d} \bms{\mu}_i$ and $\bms{\rm Q}
(R_d)=\sum_i^{\rho_i \le R_d} \bms{\rm q}_i$ denote the total dipole and
own quadrupole moment, respectively, within the sphere of radius $R_d$ and
$\bms{\rm W}(R_d)=\sum_i^{\rho_i \le R_d} \bms{\rm w}_i$. In the case of
point dipoles, we have $R_d \to R$, $\bms{\rm q}_i, \bms{\rm g}_i, \ldots
\to 0$ and the MRF (17) transforms into $\bms{M}(R)/R^3 + \bms{\rm W}(R)
\bms{\rho}/R^5$. This last formula shows that the reaction field of finite
systems is inhomogeneous even for point dipoles. Only for macroscopic ($R
\to \infty$) systems, we reproduce the well-known homogeneous reaction field
$\bms{M}(R)/R^3$ introduced by Barker and Watts [3]. For finite IS systems,
additional higher multipole terms appear. This brings, for example, into
existence of the new quadrupole-dipole and quadrupole-quadrupole interactions
in the intermolecular potential (16). We note that the idea of using the
higher multipole moments in the RF has been proposed for the first time by
Friedman [5].

However, the modified intermolecular potential (16) still needs to be
complemented by a self-consistent fluctuation formula as this has already
been done in the preceding section by the fluctuation formula (7) for the
potential of interaction in the site-site cut-off scheme (11). Unfortunately,
it is not a simple matter to construct fluctuation formulas in the molecular
cut-off approach. This problem will be considered in further studying.

The difference in the RF geometry between IS and PD models lies in the
distinction for their microscopic operators of polarization density.
For IS models
\begin{equation}
\bms{\hat P}_{\rm L}(\bms{k}) = \frac{{\rm i}\bms{k}}{k^2} \sum_{i=1}^N
{\mbox{\large e}}^{-{\rm i} \bvs{k\!\cdot\!r}_i} \sum \limits_{a} q \scs{A}
{\mbox{\large e}}^{-{\rm i} \bvs{k\!\cdot\!\delta}_i^a} =
\bms{\hat M}_{\rm L}(\bms{k})-\frac{{\rm i} \bms{k}}{2} \ \bms{\hat k}
\bms{\hat k} \,\bms{:} \sum_{i=1}^N \bms{\omega}_i {\mbox
{\large e}}^{-{\rm i} \bvs{k\!\cdot\!r}_i} + ... \ \ ,
\end{equation}
where $\bms{\hat M}_{\rm L}(\bms{k}) = \bms{\hat k} \sum_{i=1}^N \bms{\hat k}
\bms{\cdot} \bms{\mu}_i {\mbox {\large e}}^{-{\rm i} \bvs{k\!\cdot\!r}_i}$ is
the microscopic operator of polarization density for point dipoles and an
expansion over small parameter $\bms{k\!\cdot\!\delta}_i^a$ has been made
[15]. However, putting $\bms{\hat P}_{\rm L}(\bms{k}) \equiv \bms{\hat
M}_{\rm L}(\bms{k})$ in the microscopic electric field $\bms{\hat E} \scs
{R}^{\rm RF}(\bms{k})$ (6) at the very beginning and taking attempts to
perform the inverse Fourier transform, we obtain that the corresponding
integral is divergent in $\bms{k}$-space when $k \to \infty$. This divergence
is involved by the specific nature of point dipoles for which the parameter
$\bms{k\!\cdot\!\delta}_i^a$ becomes indeterminate in the limit $k \to
\infty$ because of $\bms{\delta}_i^a \to +0$ and the expansion (18) fails.
Therefore, we must manipulate with the full operator $\bms{\hat P}_{\rm L}
(\bms{k})$ to obtain the interdipolar potential (12) consequently and let
$\bms{\delta}_i^a \to +0$ at the end of the calculation only.

Since $\bms{\mu} \sim d$ and $\bms{\rm q} \sim d^2$, the quadrupole
contribution with respect to the dipole term is varied in (16) from of
order $(d/R)^2$ at $r_{ij}=0$ to $d/R$ at $r_{ij}=R_d$. Therefore, as far
as the usual intermolecular potential (8) is applied in simulations, the
dielectric constant can not be reproduced with the precision better than
$\sim d/R$. It is evident that using the modified intermolecular potential
(16) will lead to the uncertainties of order $(d/R)^2$. They decrease at
increasing the size of the sample as $R^{-2}$, i.e., with the same rate as
those connected with the truncation of the potential. Effects of the
octupole and higher order multipole contributions into the MRF
are of order $(d/R)^3$ and can be ignored.

\vspace{12pt}

\section{Applying the ISRF to a MCY water model}

\hspace{1em}  In the previous investigations [11--13], the standard PDRF
geometry (8) has been applied to actual simulations of the MCY and TIP4P
models. As a result, the static, frequency-dependent [11, 12] and
wavevector-dependent [13] dielectric constant has been determined.
For these models $d=1.837 {\rm \AA}$ and the cut-off radius $R=9.856 {\rm
\AA}$ has been used in the simulations. From the afore said in the preceding
section, it is expected that the precision of these calculations can not
exceed $d/R \sim 20 \%$. We shall show now by actual calculations that this
prediction indeed takes place.

As an example we apply the ISRF geometry (11) to the MCY potential [16].
The calculations have been performed with the help of Monte Carlo (MC)
simulations, details of which are similar to those reported earlier [13],
at the density of $\rho$= 1.0 g/cm$^3$ and at the temperature of $T=292$
K, i.e., in the same thermodynamic point and yet with the same number
$N=256$ of molecules and cut-off radius \mbox{$R=9.856 {\rm \AA}$ as
considered in [11, 13].}

Our result of the calculation (7) for the longitudinal components of the
wavevector-dependent infinite-system Kirkwood factor $g \scs{\rm L}(k)$
and dielectric constant $\varepsilon \scs{\rm L}(k)$ obtained within
the ISRF geometry is presented in Figs.~1 and 2, respectively, as the
full circles connected by the solid curves. For the purpose of comparison,
analogous calculations performed previously [13] within the PDRF are also
included in these figures (the open circles connected by the dashed curves).
It is obvious that oscillations observing in the shape of $g \scs{\rm L}
(k)$ and $\varepsilon \scs{\rm L}(k)$ obtained within the PDRF method are
nonphysical and caused by the finite molecular size which is assumed to be
zero in this approach. At the same time, the ISRF geometry gives the true,
more smooth dependencies for the Kirkwood factor and dielectric constant
because the influence of the finite molecular size is included here
explicitly. As we can see from the figures, deviations of values for the
wavevector-dependent dielectric quantities obtained using the PDRF from
those evaluated within the ISRF geometry are significant. These deviations
achieve maximal values about $25\%$ near $k=3{\rm \AA}^{-1}$, where the
Kirkwood factor has the first maximum. For great wavevector values
$(k>6{\rm \AA}^{-1})$ the both geometries lead to identical results because
the influence of boundary conditions is negligible in this range of $k$.

We remark that the wavevector-dependent quantities were calculated directly
for the discrete set $k=n k_{\rm min}$ of grid points accessible in the
simulations, where $k_{\rm min}=0.319{\rm \AA}^{-1}$ and $n$ is an integer
number. These quantities are marked in the figures by the symbols. To obtain
intermediate values between the grid points we have used the cubic spline
interpolation for the most smooth dependency, namely, for $g \scs{\rm L}(k)$.
Then values of $\varepsilon \scs{\rm L}(k)$ can be evaluated anywhere in the
considered domain of $k$-space on the basis of the interpolation values of
$g \scs{\rm L}(k)$ via Eq.~7. In particular, the first singularity of
$\varepsilon \scs{\rm L}(k)$ (see Fig.~2a) has been investigated in such
a way.

\vspace{12pt}

\section{Conclusion}

\hspace{1em}  Two alternative methods (ISRF and MRF) to overcome the
difficulties associated with finiteness of the molecule with respect
to the system size have been proposed for IS models of polar systems.
It has been shown rigorously that the fluctuation formula, which is
commonly used for the calculation of the dielectric constant in computer
experiment, corresponds to the ISRF geometry with the site-site
cut-off for Coulomb interaction potentials. The molecular cut-off
scheme leads to the MRF geometry with an additional quadrupole term
to the well-known PDRF.

It has been corroborated by actual calculations that the ISRF geometry
exhibits to be much more efficient with respect to the usual PDRF method
for the investigation of the dielectric properties of IS models. The
modified MRF approach seem to be comparable in efficiency with the ISRF
geometry. An application of the MRF to practical simulations we hope
to perform in further studying.

\vspace{2.5cm}

\begin{center}
{\large Figure captions}
\end{center}

{\bf Fig.~1.}~Longitudinal component of the wavevector-dependent
Kirkwood factor for the MCY water. The results in the ISRF and PDRF
geometries are plotted by the solid and dashed curves, respectively.

\vspace{12pt}

{\bf Fig.~2.}~Longitudinal component of the wavevector-dependent dielectric
constant for the MCY water. Notations as for fig.~1. The vertical lines
indicate positions of a singularity.


\vspace{12pt}

\begin{thebibliography}{99}
\itemsep-2pt
\small{
\bibitem  {1}  H.~Fr$\ddot{\rm o}$lich, 1959,
               {\em Theory of Dielectrics} (Clarendon Press).
\bibitem  {2}  C.J.F.~Boettcher, 1973,
               {\em Theory of Electric Polarization}, Vol.~1 (Elsevier).
\bibitem  {3}  J.A.~Barker and R.O.~Watts, Mol. Phys. 26 (1973) 789.
\bibitem  {4}  U.M.~Titulaer and J.M.~Deutch, J. Chem. Phys. 60 (1974), 1502
\bibitem  {5}  H.L.~Friedman, Mol. Phys. 29 (1975) 1533.
\bibitem  {6}  M.~Neumann and O.~Steinhauser, Mol. Phys. 39 (1980) 437.
\bibitem  {7}  M.~Neumann, O.~Steinhauser and G.S.~Pawley,
               Mol. Phys. 52 (1984) 97.
\bibitem  {8}  M.~Neumann, Mol. Phys. 57 (1986) 97.
\bibitem  {9}  I.P.~Omelyan, Phys. Lett. A 208 (1995) 237.
\bibitem {10}  I.P.~Omelyan, Mol. Phys. 87 (1996) 1273.
\bibitem {11}  M.~Neumann, J. Chem. Phys. 82 (1985) 5663.
\bibitem {12}  M.~Neumann, J. Chem. Phys. 85 (1986) 1567.
\bibitem {13}  I.P.~Omelyan, Phys. Lett. A 220 (1996) 167.
\bibitem {14}  I.P.~Omelyan, Phys. Lett. A 212 (1996) 279.
\bibitem {15}  F.O.~Raineri, H.~Resat and H.L.~Friedman,
               J. Chem. Phys. 96 (1992) 3068.
\bibitem {16}  O.~Matsuoka, E.~Clementi and M.~Yoshimine,
               J. Chem. Phys. 64 (1976) 2314.
}
\end{thebibliography}
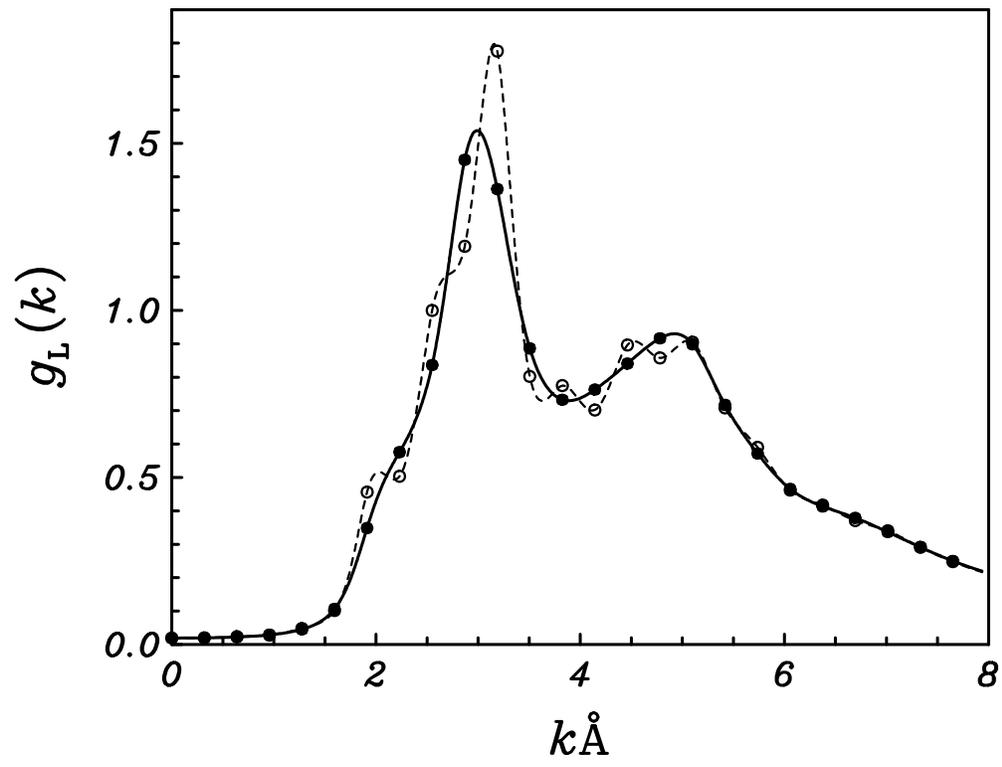
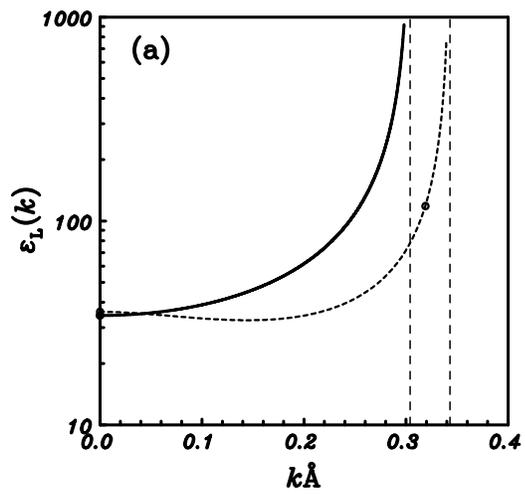
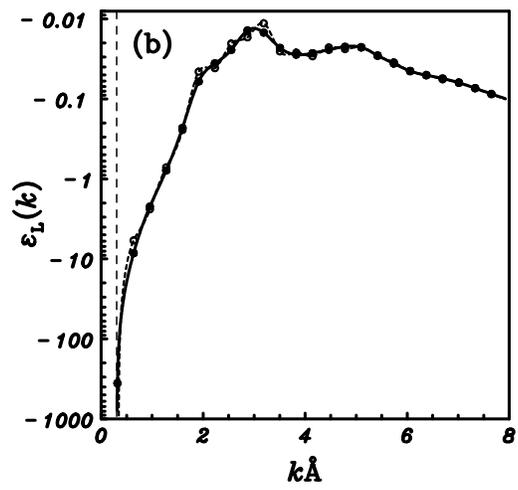
\end{document}